\documentclass{ws-ijmpa}
\def\be{\begin{equation}}
\def\beq{\begin{equation}}
\def\eeq{\end{equation}}
\def\beas{\begin{eqnarray*}}
\def\eeas{\end{eqnarray*}}
\def\ee{\end{equation}}
\def\bea{\begin{eqnarray}}
\def\eea{\end{eqnarray}}
\def\ba{\begin{array}}
\def\ea{\end{array}}

\def \da {\delta}
\def \um {\frac{1}{2}}
\def \Da {\Delta}
\def \sig {\sigma}
\def \um {\frac{1}{2}}
\def \b {\beta}
\def \a {\alpha}
\def \ga {\gamma}

\def \om {\omega}

\def \La {\Lambda}
\def\ep{\epsilon}
\def\IZ{\mathbb{Z}}
\def\IR{\mathbb{R}}
\def\IC{\mathbb{C}}

\def\II{\mathbb{I}}

\def\rref#1{(\ref{#1})}
\numberwithin{equation}{section}
\begin{document}

\markboth{J.E.NELSON, R.F.PICKEN}{A QUANTUM GOLDMAN BRACKET FOR LOOPS ON SURFACES}

%
\catchline{}{}{}{}{}
%

\title{A QUANTUM GOLDMAN BRACKET FOR LOOPS ON SURFACES}

\author {J.E.NELSON}

\address{Dipartimento di Fisica Teorica\\ Universit\`a
degli Studi di Torino and Istituto Nazionale di Fisica Nucleare, Sezione di
Torino,\\ via Pietro Giuria 1, 10125 Torino, Italy.
\\
nelson@to.infn.it}

\author{R.F.PICKEN}

\address{Departamento de Matem\'{a}tica and \\
Centro de An\'{a}lise Matem\'{a}tica, Geometria e Sistemas Din\^{a}micos (CAMGSD), \\
 Instituto Superior T\'{e}cnico, TU Lisbon\\
Avenida Rovisco Pais, 1049-001 Lisboa, Portugal.\\
rpicken@math.ist.utl.pt}

\maketitle

\begin{history}
\received{Day Month Year}
\revised{Day Month Year}
\end{history}

\begin{abstract}
In the context of (2+1)--dimensional gravity, we use holonomies of constant connections which generate a $q$--deformed representation of the fundamental group to derive signed area phases which relate the quantum matrices assigned to homotopic loops. We use these features to determine a quantum Goldman bracket (commutator) for intersecting loops on surfaces, and discuss the resulting quantum geometry.

\end{abstract}

\ccode{PACS numbers: 04.60.Kz, 02.20.Uw, Mathematics Subject Classification: 83C45}

\section{Introduction and background}	

There are many approaches to the quantization of gravity--without matter couplings--in 3 (2 space, 1 time) dimensions. We shall start with the Einstein action with nonzero cosmological constant
\beq
I_{\hbox{\scriptsize \it Ein}}
  = \int\!d^3x \sqrt{-{}^{\scriptscriptstyle(3)}\!g}\>
  ({}^{\scriptscriptstyle(3)}\!R - 2\Lambda) \label{bb1}.
\eeq
In the first order-formalism (see Refs~\refcite{Achu}--\refcite{Witt} and Refs~\refcite{NR0}--\refcite{NRZ}) this action is written as 
\beq
I_{\hbox{\scriptsize \it Ein}}
  = \int\- (d\omega^{ab}-{\omega^a}_d \wedge\omega^{db}
  +{\frac{\Lambda}{3}} e^a\wedge e^b)\wedge e^c\,\epsilon_{abc} ,
\qquad a,b,c=0,1,2.
\label{b2}
\eeq
where the triad $e^a$ is related to the metric through
\beq
g_{\mu \nu} = {e^a}_{\mu} {e^b}_{\nu} \eta_{ab} ,
\label{bc0}
\eeq
and the (2+1)-dimensional Ricci curvature and torsion are
\beq
R^{ab} = d\omega^{ab} - \omega^{ac}\wedge\omega_{c}{}^{b} ,\quad
R^{a} = de^a -\omega^{ab}\wedge e_b 
\label{bc2}
\eeq

For $\Lambda\ne0$, this action can be written (up to a total derivative) in the 
Chern-Simons form
\beq
I_{\hbox{\scriptsize CS}} = - \frac{\a}{4}
\int(d\omega^{AB}-\frac{2}{3}\omega^A{}_E\wedge\omega^{EB})
\wedge\omega^{CD} \epsilon_{ABCD} ,\qquad A,B,C,D = 0,1,2,3
\label{b3}
\eeq
with an (anti-)de Sitter spin connection $\omega^{AB}$ 

\beq
{\omega^A}_B=\left( \begin{array}{cc}
\omega^a{}_b& \frac{k e^a}{\a}\\[1ex] -\frac{e^b}{\a} & 0 \end{array} \right) .
\label{bc1}
\eeq
where the tangent space metric is $\eta_{AB}=(-1,1,1,k)$, and $k$ is the sign of $\Lambda$ 
with $\Lambda =  k\alpha^{-2}$. In Eq. \rref{b3} the Levi-Civita density is $\epsilon_{abc3}=-\epsilon_{abc}$, and 
in \rref{bc1} the triads appear as $e^a=\a\omega^{a3}$.

The corresponding curvature two-form $R^{AB}=d\omega^{AB}-\omega^{AC}\wedge\omega_C{}^B$
has components $R^{ab}+\La e^a \wedge e^b$, $R^{a3}=\frac{R^a}{\a}$, and the
field equations derived from the action \rref{b3} are simply $R^{AB}=0$, 
implying that the torsion vanishes everywhere and that the curvature 
$R^{ab}$ is constant. This can alternatively be seen from the (2+1)-dimensional splitting of spacetime, where the action \rref{b3} decomposes as
\beq
I_{\hbox{\scriptsize CS}} =
\frac{\a}{4} \int\!dt\int\!d^2x\,\epsilon^{ij}\epsilon_{ABCD}\,
 (\omega^{CD}{}_j\,{\dot{\omega}}^{AB}{}_i-\omega^{AB}{}_0 R^{CD}{}_{ij})
\label{b5}
\eeq
(with $\epsilon^{0ij} = -\epsilon^{ij}$), from which the constraints are
\beq
R^{AB}{}_{ij}=0.
\label{b8}
\eeq

The constraints \rref{b8} imply that the (anti-)de Sitter connection $\omega^{AB}{}_i$ is flat.  It can therefore be written locally in terms of an $\hbox{SO}(3,1)$ ($\Lambda > 0$) - or 
$\hbox{SO}(2,2)$ ($\Lambda < 0$) - valued zero-form $\psi^{AB}$ as $d\psi^{AB}=\om^{AC}\, 
\psi_C{}^B$. It is actually more convenient to use the spinor groups 
$\hbox{SL}(2,\IR)\otimes \hbox{SL}(2,\IR)$ (for $\hbox{SO}(2,2)$)
and $\hbox{SL}(2,\IC)$ (for $\hbox{SO}(3,1)$).  (Details of the spinor
group decomposition can be found in Ref.~\refcite{NRZ}.)  Define the one-form
\beq
\Da(x) = \Da_{i}(x)dx^{i} = \frac{1}{4}\omega^{AB}(x)\ga_{AB}
\label{bc4}
\eeq
where $\ga_{AB} = \um \left[\ga_A,\ga_B \right]$ and the $\ga_A$ are Dirac matrices. Eq. \rref{b8} now implies that $d\Da-\Da\wedge\Da=0$. The corresponding local or "pure gauge"
expression for $\Da$ is
\beq
dS(x)=\Da(x)S(x) .
\label{bc5}
\eeq
where $S$ are multivalued $\hbox{SL}(2,\IR)$ or $\hbox{SL}(2,\IC)$ matrices. 

The above discussion means that the $\hbox{SO}(3,1)$ or $\hbox{SO}(2,2)$ - valued $\psi^{AB}$, or the $\hbox{SL}(2,\IR)$ or $\hbox{SL}(2,\IC)$ matrices $S$ can be interpreted as holonomies, when the connections $\om^{AB}$ or $\Da$ are integrated along closed paths (loops) on the two--dimensional surface $\Sigma$. The flatness of the connection $\Delta$ implies that each $S[\gamma]$ depends only on the homotopy class of $\gamma$. Further, the matrices $S$ are not gauge invariant, but are gauge covariant i.e. under a gauge transformation (a change of base point), they transform by conjugation.  

The Einstein action can be used to gain further information about these holonomies. For example, the Poisson brackets of the $\om^{AB}$ can be read off from \rref{b5}:
on a $t=\hbox{const.}$ surface $\Sigma$,
\beq
\{\omega^{AB}{}_{i}(x),\omega^{CD}{}_{j}(y)\}
 =\frac{k}{2\a}\epsilon_{ij} \epsilon^{ABCD} \da^{2}(x-y).
\label{b6}
\eeq
and the spinor version is 
\begin{eqnarray}
\{\Da_{i}^{\pm}(x), \Da_{j}^{\pm}(y)\}
 &=& \pm \frac{i}{2\a\sqrt{k}}\ep_{ij}\sigma^{m}\otimes\sigma^{m}\da^{2}(x-y)
 \nonumber\\
\{\Da_{i}^{+}(x), \Da_{j}^{-}(y)\} &=& 0 ,
\label {d1}
\end{eqnarray}
where the $\sigma^{m}$ are Pauli matrices, the $\pm$ refer to the
decomposition of the $4\times 4$ representations of $\Da(x)$, $S(x)$ into
$2\times 2$ irreducible parts (see Ref.~\refcite{NRZ}) and $\sqrt k$ means $+1$ for $k=1$ and $+i$ for $k=-1$.

The Poisson brackets Eqs. \rref{b6} and \rref{d1}, when integrated along loops $\ga, \sig$ 
(with $\ga,\sig\!\in\!\pi_{1}(\Sigma,x_0)$) yield the Poisson brackets of the components of the holonomies 
$\psi^{AB}$ 
\beq
\{{\psi^{AB}}_{\ga},  {\psi^{CD}}_{\sig}\} = -k \epsilon^{ABCD}.
\label{p1}
\eeq
and a similar (complicated) expression for the $S^\pm$. The matrices $S^{\pm}[\gamma]$ thus 
furnish a representation of $\pi_{1}(\Sigma,x_0)$ in $\hbox{SL}(2,\IR)$ or
$\hbox{SL}(2,\IC)$.  Under a gauge transformation the $S^\pm$ transform by conjugation, so their traces provide an (overcomplete) set of gauge-invariant Wilson loop variables.

The classical Poisson brackets for these trace variables were calculated by
hand for the genus $1$ and genus $2$ cases, and then generalized and
quantized in Ref.~\refcite{NR1}.  A closely related quantum algebra was calculated 
in Ref.~\refcite{ch} using the technique of "fat graphs". The classical Poisson 
bracket algebra also appears (see Ref. \refcite{ug}) in the study of Stokes matrices 
(monodromy data) which relate the solutions of matrix differential equations. For genus 1 the Poisson algebra is
\beq
\{R_1^{\pm},R_2^{\pm}\}=\mp\frac{i}{4\a\sqrt k}(R_{12}^{\pm}-
 R_1^{\pm}R_2^{\pm}) \quad \hbox{\it and cyclical permutations},
\label{b7}
\eeq
where $R^{\pm}= \um \hbox{Tr}S^{\pm}$.  Here the subscripts $1$ and $2$
refer to the two independent intersecting circumferences $\gamma_1$, $\gamma_2$
on $\Sigma$ with intersection number $+1$,\footnote{Paths with intersection
number 0, $\pm$ 1 are sufficient to characterize the holonomy algebra for
genus $1$.  For $g>1$, one must in general consider paths with two or more
intersections, for which the brackets \rref{b7} are more complicated; see
Refs.~\refcite{NR3,NR4}.} while the third traced holonomy, $R^\pm_{12}$, corresponds to
the path $\gamma_1\cdot\gamma_2$, which has intersection number $-1$ with
$\gamma_1$ and $+1$ with $\gamma_2$.  

Classically, the six traced holonomies $R^\pm_{1,2,12}$ provide an overcomplete
description of the spacetime geometry of $\IR\!\times\!T^2$. Consider the cubic polynomials
\begin{eqnarray}
F^{\pm} &=& 1-(R_1^{\pm})^2-(R_2^{\pm})^2-(R_{12}^{\pm})^2 +
 2 R_1^{\pm}R_2^{\pm}R_{12}^{\pm} \nonumber\\
        &=& \um\, \hbox{Tr}\left(I -
 S^{\pm}[\ga_1]S^{\pm}[\ga_2]S^{\pm}[\ga_1^{-1}]S^{\pm}[\ga_2^{-1}]\right),
\label{b9}
\end{eqnarray}
where the last equality follows from the identities
$$
A + A^{-1} = I\,\hbox{Tr}A
$$
for $2 \times 2$ matrices $A$ with determinant $1$. These polynomials have vanishing Poisson 
brackets with all of the traces $R_a^{\pm}$, and are cyclically symmetric in the 
$R_a^{\pm}$. The $F^\pm$ vanish classically by
the $\hbox{SL}(2,\IR)$ or $\hbox{SL}(2,\IC)$ Mandelstam identities,
which can be viewed as the application of the fundamental relation of $\pi_{1}$ of the torus
\be
\ga_1^{\vphantom{-1}}\cdot\ga_2^{\vphantom{-1}}\cdot\ga_1^{-1}\cdot\ga_2^{-1}
  = {\II}
\label{gp}
\ee
to the representations $S^{\pm}$ occuring in the last line of \rref{b9}.

In this approach, the constraints have been solved exactly. There is no Hamiltonian, and 
no time development. This formalism describes either initial data for some (unspecified) choice of time, or the time-independent spacetime geometry.

We can quantize the classical algebra \rref{b7} by firstly replacing the classical Poisson brackets  $\{\,,\,\}$ with commutators $[\,,\,]$, with the rule
\beq
[x,y]= xy-yx =i \hbar \{x,y\} ;
\eeq
and secondly, on the right hand side (r.h.s.) of \rref{b7}, replacing the product with
the symmetrized product,
\beq
xy \to \um (xy +yx) .
\eeq
The resulting operator algebra is given by
\beq
\hat R_1^{\pm}\hat R_2^{\pm}e^{\pm i \theta}
  - \hat R_2^{\pm}\hat R_1^{\pm} e^{\mp i \theta}=
  \pm 2i\sin\theta\, \hat R_{12}^{\pm} \quad \hbox{\it and cyclical
  permutations}
\label{za}
\eeq
with $\tan\theta= {i \sqrt k\hbar} /{8\a}$.
Note that for $\Lambda\!<\!0$, $k=-1$, and $\theta$ is real, while for
$\Lambda\!>\!0$, $k=1$, and $\theta$ is pure imaginary.

The algebra \rref{za} is not a Lie algebra, but it is related to the
Lie algebra of the quantum group $\hbox{SU}(2)_q$ Refs. \refcite{NRZ,su}, where
$q=\exp{4i\theta}$, and where the cyclically invariant $q$-Casimir is
the quantum analog of the cubic polynomial \rref{b9},
\beq
\hat F^{\pm}(\theta)
  = {\cos}^2\theta- e^{\pm 2i\theta} \left( (\hat R_1^\pm)^2+
(\hat R_{12}^\pm)^2\right) -e^{\mp 2i\theta} (\hat R_2^\pm)^2
 + 2e^{\pm i\theta}\cos\theta \hat R_1^\pm \hat R_2^\pm \hat R_{12}^\pm .
\eeq

For $g>1$ it was shown in Ref.~\refcite{gav} that the algebra calculated in Ref.~\refcite{NR1} is isomorphic to a non--standard deformation of $\hbox{SO}(2g+2)_q$.

The representations of the algebra \rref{za} have been studied e.g. in Ref.~\refcite{NRZ}. Here we choose to represent each $(\pm)$ copy of the $\hat R_a^{\pm}$ as
\beq
\hat R_a = \um (\hat A_a + \hat A_a^{-1})
\eeq
where, from \rref{za} the $\hat A_a$ must satisfy (here we discuss the $(+)$ algebra, the $(-)$ algebra has $q^{-1}$ rather than $q$)
\beq
\hat A_1 \hat A_2 = q \hat A_2 \hat A_1 \quad \hbox{\it and cyclical permutations}
\label{fund}
\eeq
Relations of the type \rref{fund} are called quantum plane relations, or $q$--commutators, or  Weyl pair relations. Returning to the {\it untraced matrices} $S$ one notes that, writing them  in diagonal form as
\beq
S(\gamma_i) = U_i = \left(\begin{array}{clcr}A_i & 0\\0 & A_i^{-1}\end{array}\right) 
\quad i=1,2 
\eeq
it follows that the {(now \it quantum matrices}) $\hat U_1, \hat U_2$ must satisfy {\it by both matrix and operator multiplication}, the $q$--commutation relation
\beq
\hat U_1 \hat U_2 = q \hat U_2 \hat U_1 
\label{fund2}
\eeq
i.e. they form a matrix--valued Weyl pair. Equation \rref{fund2} can be understood as a deformation of Eq. \rref{gp}. 
 
The present authors decided to study the quantum matrices $\hat U_1, \hat U_2$ which satisfy \rref{fund2}. Consider the diagonal representation 
\be
\hat U_i = \left(\begin{array}{clcr}e^{{\hat r_i}}& 0 \\0& e^{-{\hat r_i}}\end{array}\right) = e^{{\hat r_i}\sigma_3}
\quad i=1,2
\label{hol}
\ee
where $\sigma_3$ is a Pauli matrix. From the identity
\be e^{\hat X} e^{\hat Y}= e^{\hat Y} e^{\hat X} e^{[ \hat X, \hat Y ]},
\label{bch}
\ee
valid when $[ \hat X, \hat Y ]$ is a $c$--number, it follows that the quantum parameters $\hat r_1, \hat r_2$ (also used in Ref.~\refcite{cn}) satisfy the commutator
\beq
[\hat r_1, \hat r_2] = - \frac{i\hbar \sqrt{-\Lambda}}{4}.
\label{comm}
\eeq

We note that in order to make the connection with $2+1$--dimensional gravity it is necessary to consider {\it both} $\hbox{SL}(2,\IR)$ sectors. The mathematical properties of just one sector have been studied in Ref.\refcite{NP3}. In Section \ref{qmp} a brief review of quantum matrices is given, whereas Section \ref{hom} discusses quantum holonomy matrices for homotopic paths, and shows how they are related by the signed area between the two paths. Section \ref{go} uses these concepts to quantize a classical bracket due to Goldman (Ref.~\refcite{gol}), thus obtaining commutators between intersecting loops on surfaces.

\section{Quantum Matrix Pairs\label{qmp}}

Quantum matrix pairs - namely the quantum matrices $\hat U_1, \hat U_2$  which satisfy \rref{fund2} may, as mathematical objects, be thought of as a simultaneous  
generalization of two familiar notions of ``quantum mathematics'',  namely the quantum  
plane and quantum groups. Briefly, the quantum plane is described by two 
non-commuting coordinates $x$ and $y$ satisfying the relation 
\be 
x y = q y x 
\label{qplane} 
\ee 
whereas, for an example of a quantum group, consider the  $2\times 2$ matrices of the form  
\be
U=\left( \begin{array}{cc}a &b\\ c&d\end{array} \right) 
\label{qmatrix} 
\ee
with non-commuting entries satisfying  
\bea 
ab=qba; \quad &ac=qca;\quad &ad-da=(q-q^{-1})bc; \nonumber\\  
      bc=cb; \quad &bd=qdb; \quad &cd=qdc. 
\label{qgrel} 
\eea
A good description of matrices of the type \rref{qmatrix} whose entries satisfy \rref{qgrel} can be found in Ref. \refcite{Vokos}, but for our purposes maybe the most important property is that the matrix $U^n$ is another matrix {\it of the same type} with $q$ substituted by $q^n$.

These two concepts - the quantum plane and quantum groups - are not unrelated. Consider 
the column vector whose entries are the non-commuting coordinates $x$ and $y$. It can be checked that the components of the new column vector
\be 
\left(\begin{array}{c}{x^{\prime}}\\{y^{\prime}}\end{array}\right) = U \left(\begin{array}{c}{x}\\{y}\end{array}\right)
\ee
also satisfy 
\be 
x^{\prime} y^{\prime} = q y^{\prime} x^{\prime}  
\ee

Quantum matrices in both the diagonal and upper--triangular sectors satisfying
the fundamental relation \rref{fund2} have been studied in Refs. \refcite{NP1,NP2}. They combine the preservation of internal relations under multiplication, a quantum-group-like feature, with the fundamental $q$-commutation relation which holds between the two matrices.  The non-trivial internal commutation relations arose in the following way: in the upper-triangular sector, it was found that trivial internal commutation relations for each matrix were not compatible with the fundamental relation \rref{fund}, in that the resulting products no longer had commuting entries. However it was possible to determine patterns of non-trivial internal relations which are preserved under matrix multiplication.

For example, consider the pair of matrices satisfying \rref{fund2}
\be
U_i=\left( \begin{array}{cc} \a_i &\b_i\\ 0&{\a_i}^{-1}\end{array}
\right), \,\,\,\,i=1,2
\label{Ui}
\ee
It can be checked that, apart from the mutual relations
\be
 \a_1\a_2=q\a_2\a_1, \quad   \a_1\b_2=q\b_2{\a_1}^{-1}, \quad  \a_2\b_1=q^{-1}\b_1{\a_2}^{-1}. 
\label{mut}
\ee
which guarantee \rref{fund2}, the entries must also satisfy the following internal relations 
\be
 \a_i\b_i=\b_i{\a_i}^{-1},\quad i=1,2.
\label{int}
\ee

In Ref.~\refcite{NP2} it was shown that indeed products of powers of these matrices have the same structure of internal relations, and also taking two different products gives rise to new quantum matrix pairs of the same type. However, the internal relations \rref{int} differ in structure from the relations \rref{qgrel}, and, moreover, do not simplify in the limit $q\rightarrow 1$, which distinguishes them from e.g. Majid's braided matrices (see Ref. \refcite{maj}). 

\section{Homotopy and signed area\label{hom}} 

Consider quantum holonomy matrices simultaneously conjugated into diagonal form 
(conjugating both matrices by the same matrix $S\in SL(2,\IR)$ ) (see Eq. \rref{hol} which for convenience is repeated here) 
$$
\hat U_i=\left(\begin{array}{clcr}e^{{\hat r_i}}& 0 \\0& e^{- {\hat r_i}}
\end{array}\right), \quad i=1,2. 
$$
where the $\hat r_i, i = 1,2$ satisfy \rref{comm}. They can be thought of as arising from constant connections $\hat A$ as in Ref.~\refcite{mik}
\be
\hat U_i = \exp \int_{\ga_i} \hat A, \quad
\hat A = (\hat r_1 dx + \hat r_2 dy)
\left(\begin{array}{clcr} 1& 0 \\0& -1 \end{array}\right).
\label{hol1}
\ee
where $x,y$ are coordinates, with period 1, on the torus 
$T^2= {\IR^2}_{(x,y)}/\IZ^2$, and $y$ is constant along
$\ga_1$ and $x$ is  constant along $\ga_2$.

We have investigated constant matrix--valued connections which
generalize the connections \rref{hol1}, and applied them to a much larger
class of loops, extending the assignments $\ga_1\mapsto U_1,\,\ga_2 \mapsto U_2$ by using 
the quantum connection \rref{hol1} in the diagonal case. The larger
class of loops are represented by piecewise linear (PL) paths between integer
points in $\IR^2$, using a representation of $T^2$ as $\IR^2/\IZ^2$. We show
that the matrices for homotopic paths are related by a phase expressed in terms
of the signed area between the paths. This leads to a definition of a $q$--deformed 
representation of the fundamental group where signed area phases relate the quantum matrices
assigned to homotopic loops. 

Consider piecewise-linear (PL) paths on the plane $\IR^2$ starting at the
origin $(0,0)$  and ending at an integer point $(m,n), \, m,n\in \IZ$. Under the
identification $T^2=\IR^2/\IZ^2$, these paths give rise to closed loops on
$T^2$. The integers $m$ and $n$ are the winding numbers of the loop in the
$\ga_1$ and  $\ga_2$ directions respectively, and two loops on $T^2$ are
homotopic to each other if and only if the corresponding paths in $\IR^2$ end
at the same point $(m,n)$ 

Suppose a PL path $p$ consists of $N$ straight segments $p_1, \dots, p_N$. Any
such segment $p_i$ may be translated to start at the origin and end at
$(m,n)\in \IR^2$  (here we use the fact that the connection $A$ is invariant
under spatial translations). Then we assign to each segment $p_i$ the quantum 
matrix

\be
U_{(m,n)} = \exp \int_{p_i} A = \exp \left((mr_1 + nr_2)\sigma_3 \right) = 
\left(\begin{array}{cc} e^{mr_1+nr_2}& 0 \\0& e^{-mr_1-nr_2} 
\end{array}\right)\label{phi}   
\ee
where $\sigma_3 = \left(\begin{array}{clcr} 1& 0 \\0& -1 
\end{array}\right)$,  
and to the path $p$ the product matrix
\be
p\mapsto U_p := \prod _{i=1}^N \exp \int_{p_i} A.
\label{Up}
\ee

This assignment is obviously multiplicative under multiplication of
paths, $(p,p')\mapsto p\circ p'$, which corresponds to translating $p'$ to
start at the endpoint of $p$ and concatenating.

Now consider the {\it straight} path from $(0,0)$ to $(m,n)$. For example, with 
$U_1 = U_{(1,0)}$, $U_2 = U_{(0,1)}$ these correctly obey the fundamental relation \rref{fund2}, which can be generalized to arbitrary straight paths, using Eq. \rref{bch}.

\be
U_{(m,n)} U_{(s,t)} = q^{mt-ns} U_{(s,t)} U_{(m,n)},
\label{uu1}
\ee
where $U_{(m,n)}$  is given by Eq. \rref{phi}.

Equation \rref{uu1} expresses the relation between the quantum matrices assigned 
to the two paths going from $(0,0)$ to $(m+s,n+t)$ in two different ways around 
the parallelogram generated by $(m,n)$ and $(s,t)$,  It is also straightforward to show a triangle equation
\be
U_{(m,n)} U_{(s,t)} = q^{(mt-ns)/2} U_{(m+s,n+t)},
\label{tri}
\ee
which can be derived from the identity
$$ e^{\hat X} e^{\hat Y}= e^{\hat X + \hat Y} e^{\frac{[\hat X, \hat Y ]}{2}},
$$
which follows from \rref{bch}. 

Note that in both cases the exponent of $q$ relating the two homotopic paths
is equal to the {\em signed area} between the path $p$ on the 
left hand side (l.h.s.) and the path $p'$ on the r.h.s. 
 i.e. equal to the area between $p$ and $p'$, when the PL loop
consisting of $p$ followed by the inverse of $p'$ is oriented anticlockwise,
and equal to minus the area between $p$ and $p'$, when it is oriented
clockwise. The signed area for the parallelogram is
given by 
$\det 
\left(\begin{array}{cc} m& s\\n& t 
\end{array}\right)
=mt-ns$
and for the triangle by $\um (mt-ns)$. 

The discussion can be generalized to arbitrary non-self-intersecting PL
paths $p$ and $p'$ which connect $(0,0)$ to the same integer point $(m,n)$ in
$\IR^2$. These two paths may intersect each other several times, either
transversally, or when they coincide along a shared segment. Together they
bound a finite number of finite regions in the $xy$-plane. Now choose a
triangulation of a compact region of $\IR^2$ containing and compatible with the
paths $p,\,p'$, in the sense that each segment of the paths is made up of one
or more edges of the triangulation. We take all the triangles in the
triangulation to be positively oriented in the sense that their boundary is
oriented anticlockwise in $\IR^2$. Since $p$ and $p'$ are  homotopic, they are
homologous, and because $H_3$ of the plane is trivial, there is a unique
$2$-chain $c(p,p')$ such that $\partial c(p,p')=p-p'$. Let this chain be given
by 
\be
c(p,p') = \sum_{\alpha \in R}n_\alpha t_\alpha,
\label{2chain}
\ee
where $t_\alpha$ is a triangle of the triangulation indexed by $\alpha$ in the
index set $R$, and  $n_\alpha\ = \pm 1$ or $0$. Note that only triangles from
the finite  regions enclosed by $p$ and $p'$ can belong to the support of the 
$2$-chain, and that the coefficient of any two triangles in the same finite region
is the same. 

The {\em signed area} between $p$ and $p'$ is
\be
S(p,p') = \sum_{\alpha \in R} n_\alpha A(t_\alpha),
\label{spp}\ee
where 
$A(t_\alpha)$ is the area of the triangle $t_\alpha$. 
This is clearly independent of the
choice of  triangulation of $\IR^2$ compatible with $p,\,p'$, since the sum
of the areas of the  triangles inside each enclosed region is the area of
that region, whatever the triangulation. It follows that
\be
 U_p = q^{S(p,p')}U_{p'}.
\ee
 

\section{Goldman bracket\label{go}}

There is a classical bracket due to Goldman Ref.~\refcite{gol} for functions $T(\ga)={\rm tr}\, U_\ga$ defined on homotopy classes of loops $\ga$, which for $U_\ga \in SL(2,\IR)$ is:
\be
\{T(\ga_1), T(\ga_2)\} = \sum_{S \in \ga_1 \sharp \ga_2}
\epsilon(\ga_1,\ga_2,S)(T(\ga_1S\ga_2) -
T(\ga_1S\ga_2^{-1})).
\label{gold}
\ee

Here $\ga_1 \sharp \ga_2$ denotes the set of (transversal) intersection points
of $\ga_1$ and $\ga_2$ and $\epsilon(\ga_1,\ga_2,S)$ is the intersection index
for the intersection point $S$.  $\ga_1S\ga_2$ and $\ga_1S\ga_2^{-1}$ denote
loops which are {\it rerouted} at the intersection point $S$. In the
following we show how Eq.~\rref{gold} may be quantized using the concept of area phases 
for homotopic paths which was outlined in Section \ref{hom}. 


In order to study intersections of ``straight'' loops, represented in $\IR^2$ by straight paths
between $(0,0)$ and integer points $(m,n)$, consider their reduction to a fundamental 
domain of $\IR^2$, namely the square with vertices
$(0,0), (1,0), (1,1), (0,1)$.

\begin{figure}[hbtp]
\centering
\includegraphics[height=2cm]{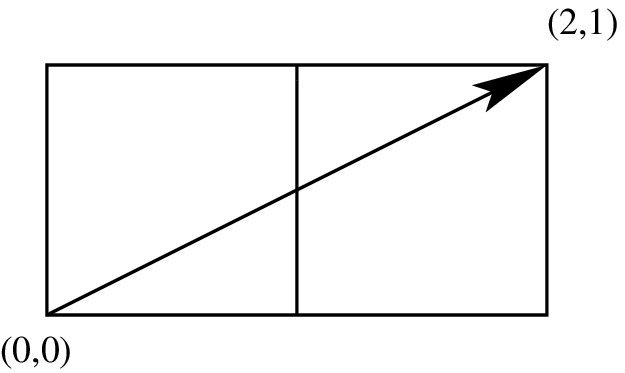}
\hspace{2cm}
\includegraphics[height=2cm]{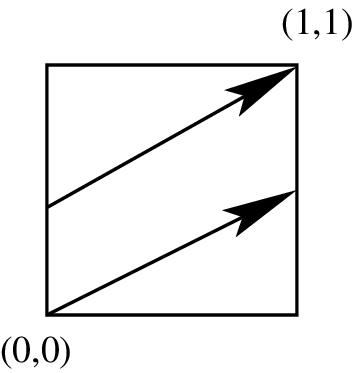}
\caption{The path $(2,1)$ and its fundamental reduction}
\label{21}
\end{figure}

Here are two examples of fundamental reduction. Figure \ref{21} shows a path
in the first quadrant, namely the path $(2,1)$, and its reduction to the
fundamental domain

\begin{figure}[hbtp]
\centering
\includegraphics[width=2cm]{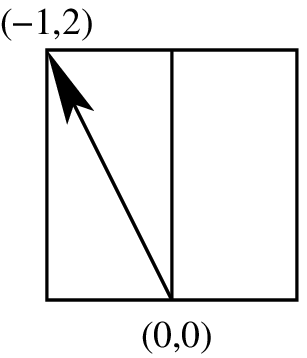}
\hspace{2cm}
\includegraphics[width=2cm]{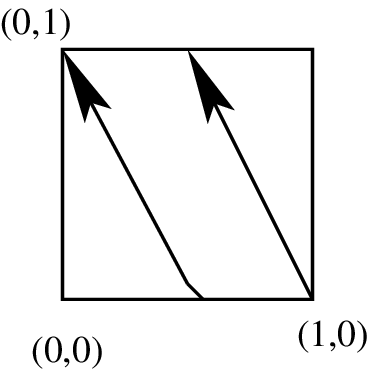}
\caption{The path $(-1,2)$ and its fundamental reduction}
\label{-12}
\end{figure}

\noindent whereas in other quadrants fundamentally reduced paths start at 
other vertices (not $(0,0)$). For example in the second quadrant the path 
$(-1,2)$ will (in the fundamental domain) start at $(1,0)$ and end at $(0,1)$, 
as shown in Figure \ref{-12}.  

When the path $(m,n)$ is a multiple of another integer path, we say it is
{\it reducible}. Otherwise it is irreducible. 

It should be clear that two paths intersect at points where their fundamental reductions intersect. We may only consider transversal intersections, namely when their respective tangent vectors are not collinear. For intersecting paths of multiplicity $1$, their
intersection number at that point is $+1$ if the angle from the first tangent
vector to the second is between $0$ and $180$ degrees, and $-1$ if between
$180$ and $360$ degrees. For paths of multiplicity greater than $1$, the
intersection number is multiplied by the multiplicities of the paths involved.
Denote the intersection number between two paths $p_1$ and $p_2$ at
$P$ (or $P, Q, R$ if more than one) by $\epsilon(p_1,p_2,P)$. The total
intersection number for two paths is the sum of the intersection numbers for
all the intersection points, denoted  $\epsilon(p_1,p_2)$. Here are three simple
(and not so simple) examples (in the fundamental domain) of single and multiple
intersections.  
\begin{enumerate}
\item If $p_1= (1,0)$ and $p_2=(0,1)$ there is a single intersection\\ 
at $(0,0)$ with $\ep=+1$.

\hfill\includegraphics[width=2cm]{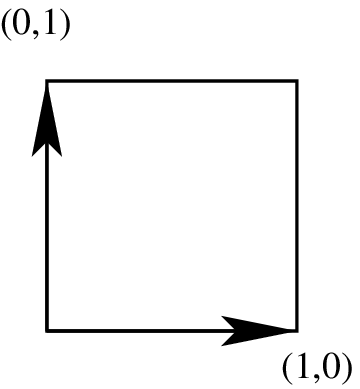}

\item If $p_1= (2,1)$ and $p_2=(0,1)$ there are two intersections,\\
at $P=(0,0)$ and $Q=(0,\um)$, each with $\ep=+1$. The total \\
intersection number is $\ep=+2$

\hfill\includegraphics[width=2cm]{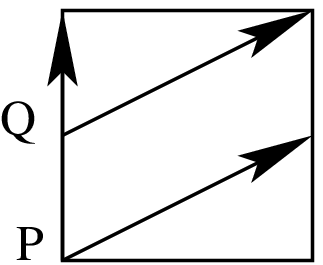}
 
\item If $p_1= (1,2)$ and $p_2=(2,1)$ there are three intersections,\\
at $P=(0,0)$, $Q=(\frac{2}{3},\frac{1}{3})$ and $R(\frac{1}{3},\frac{2}{3})$ (see figure),
each with\\ $\ep=-1$. The total intersection number is $\ep=-3$ (the point\\
 $S=(1,1)$ does not contribute since it coincides with the point $P$

\hfill\includegraphics[width=2cm]{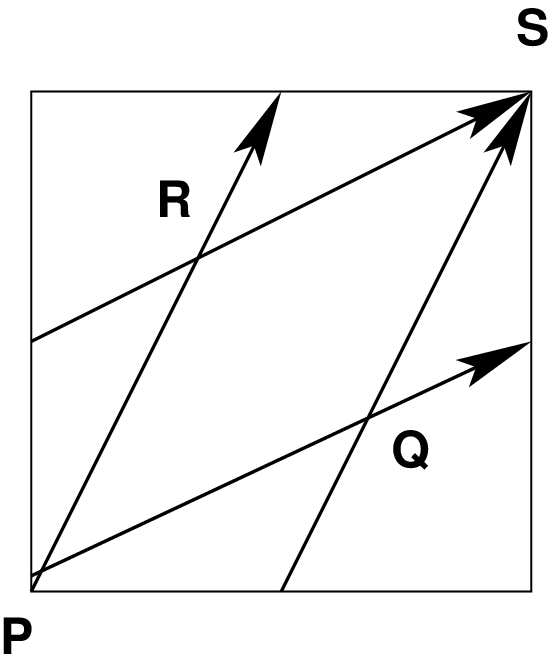}

\end{enumerate}
It should be noted that 
\begin{enumerate}
\item all intersections between a given pair of straight paths have the same
sign, since in this representation their tangent vectors have constant
direction along the loops. 

\item the total intersection number between $p_1=(m,n)$ and $p_2=(s,t)$ 
is the determinant 
\be
\ep(p_1,p_2) =  \left |\ba{clcr}m&n\\s&t\ea
\right | = mt - ns
\label{detint}
\ee
since the total intersection number is invariant under deformation, i.e. homotopy
\begin{eqnarray}
\ep((m,n),(s,t)) & = & \ep((m,0)+(0,n), (s,0)+(0,t)) \nonumber \\
& = & \ep((m,0),(0,t)) + \ep((0,n),(s,0)) \nonumber \\
& = & mt-ns.
\end{eqnarray}

\leftline{Relation \rref{detint} is easily checked for the above examples.}
\end{enumerate}

Now consider two straight paths $p_1$ and $p_2$ intersecting at the point $P$. Their positive 
and negative reroutings are denoted $p_1Pp_2$ and  $p_1Pp_2^{-1}$ respectively, where
$p_2^{-1}= (-s,-t)$ if  $p_2=(s,t)$. These reroutings are defined as
follows: starting at the basepoint follow $p_1$ to $P$, continue on $p_2$
(or $p_2^{-1}$) back to $P$, then finish along $p_1$.  Note that, in
accordance with the above rule, if the intersection point $P$ is the basepoint
itself, the reroutings $p_1Pp_2$ and  $p_1Pp_2^{-1}$ start by following
$p_2$ (or $p_2^{-1}$) from the basepoint back to itself, and then
follow $p_1$ from the basepoint back to itself. Here we show the
reroutings ($p_1Pp_2$ and $p_1Pp_2^{-1}$ respectively, and at the
various intersection points $P, Q, R$ if more than one) for the three previous examples, using the non--reduced paths which here are more convenient. 

\begin{enumerate}

\item $P=(0,0), p_1 = (1,0), p_2 = (0,1)$ 
\hfill
\includegraphics[width=2cm]{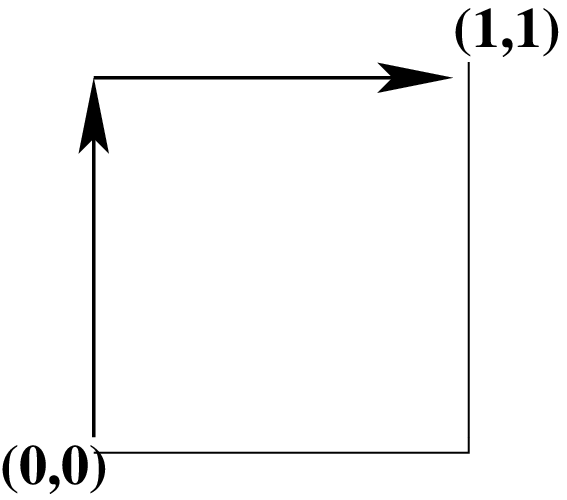}
\hspace{0.5cm}
\includegraphics[width=2.2cm]{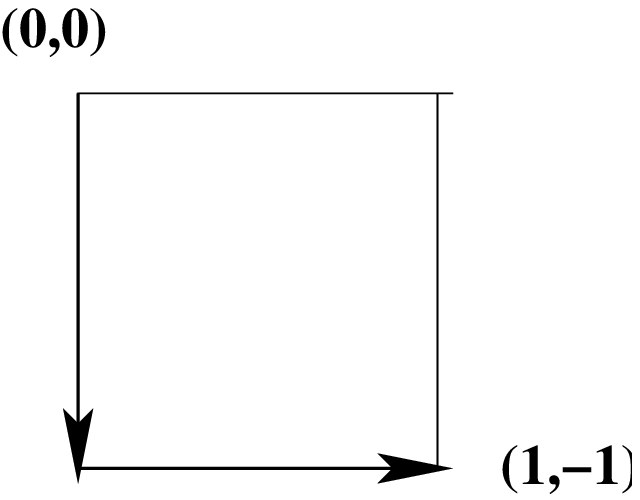}

\item (a)~$P=(0,0), p_1 = (2,1), p_2 = (0,1)$
\hfill
\includegraphics[width=2cm]{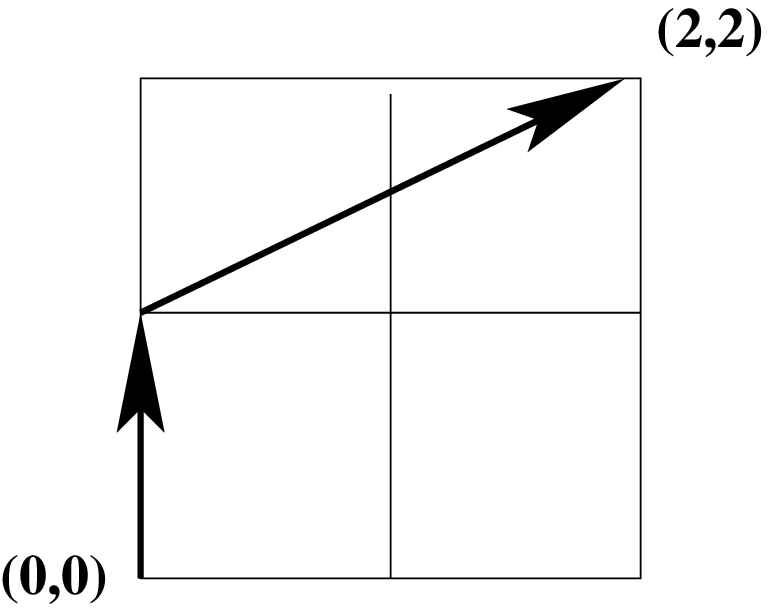}
\hspace{0.5cm}
\includegraphics[width=2cm]{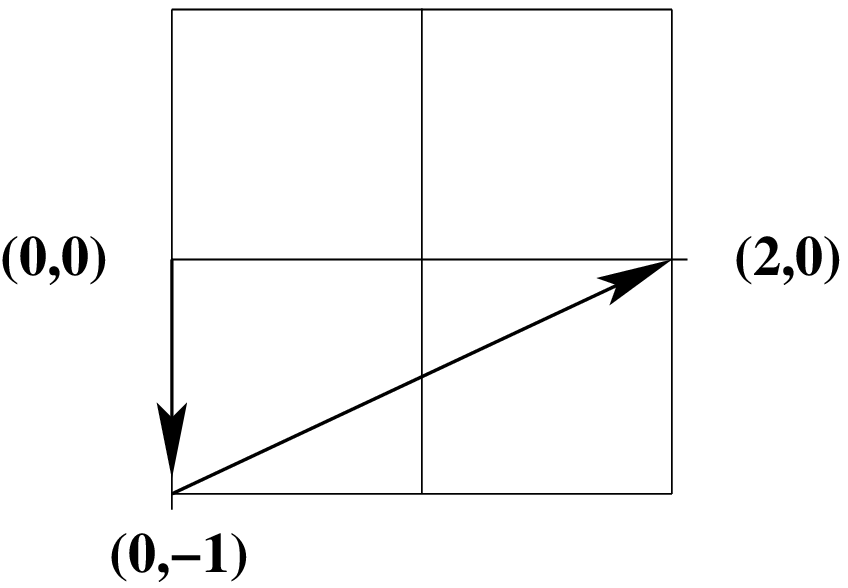}
\vspace{.5cm}
\noindent (b)~$Q=(0,\um), p_1 = (2,1), p_2 = (0,1)$
\hfill
\includegraphics[width=2cm]{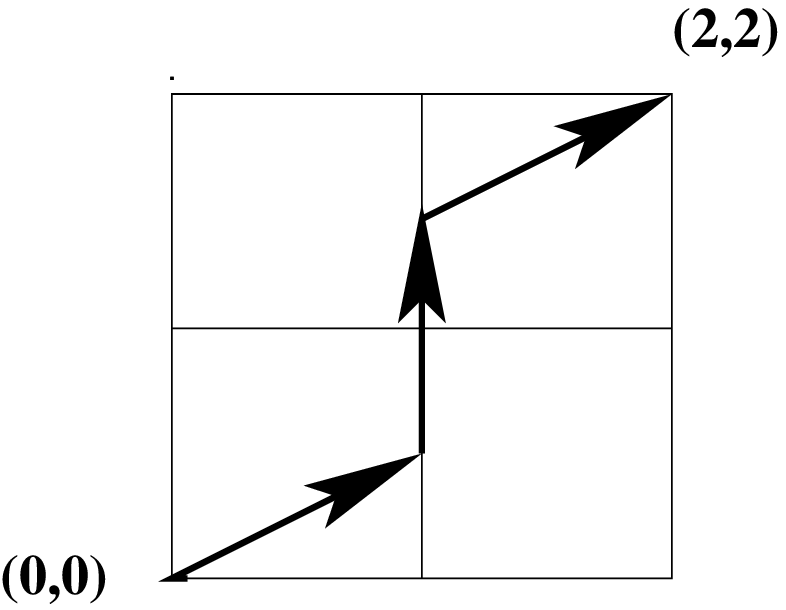}
\hspace{0.5cm}
\includegraphics[width=2cm]{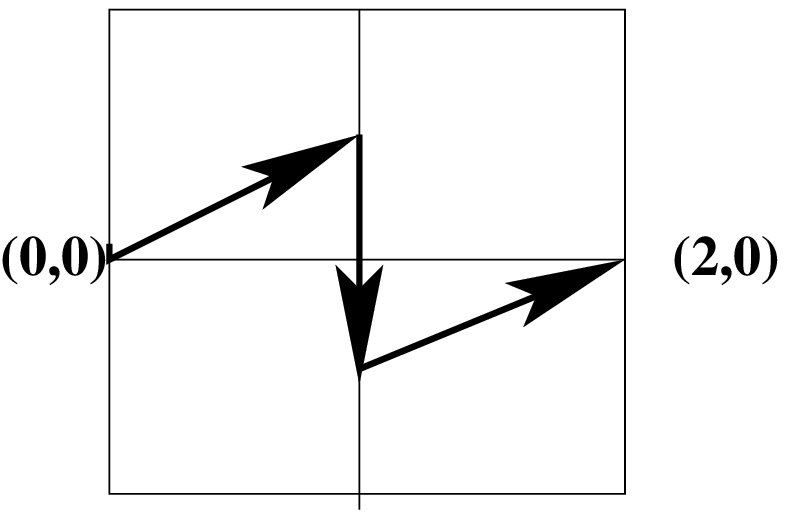}

\item (a)~$P=(0,0), p_1 = (1,2), p_2 = (2,1)$
\hfill
\includegraphics[width=2.5cm]{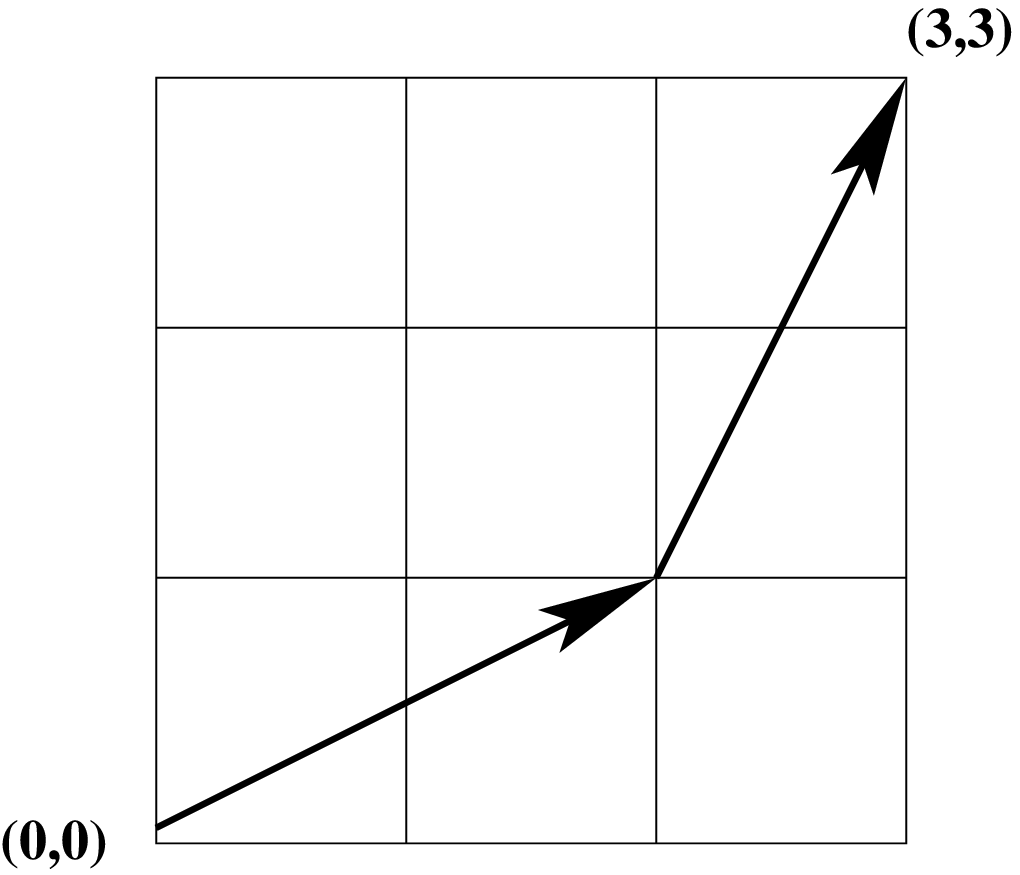}
\hspace{0.5cm}
\includegraphics[width=2cm]{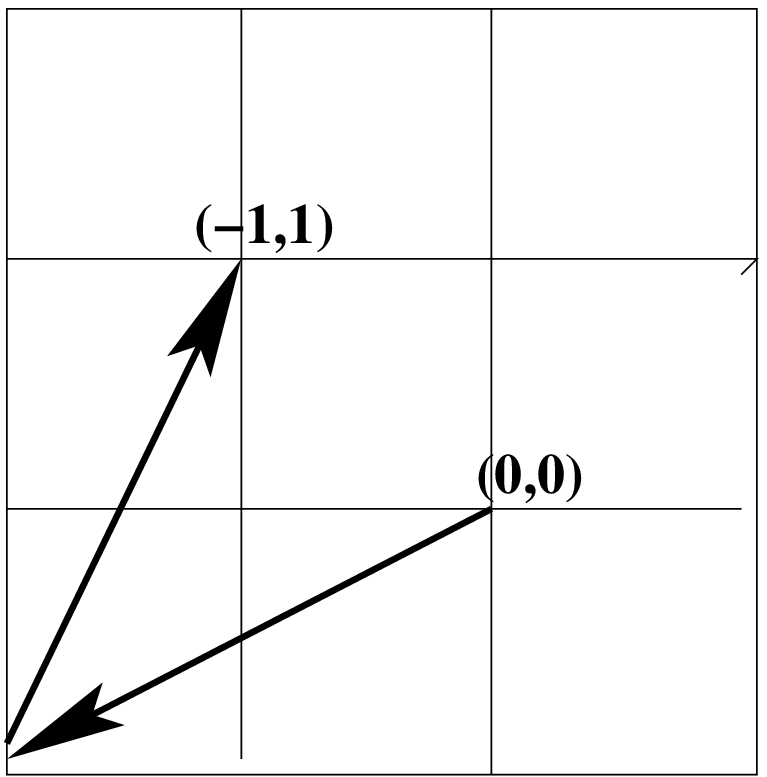}
\vspace{.5cm}
\noindent (b)~$Q=(\frac{2}{3},\frac{1}{3}), p_1 = (1,2), p_2 = (2,1)$
\hfill
\includegraphics[width=2.5cm]{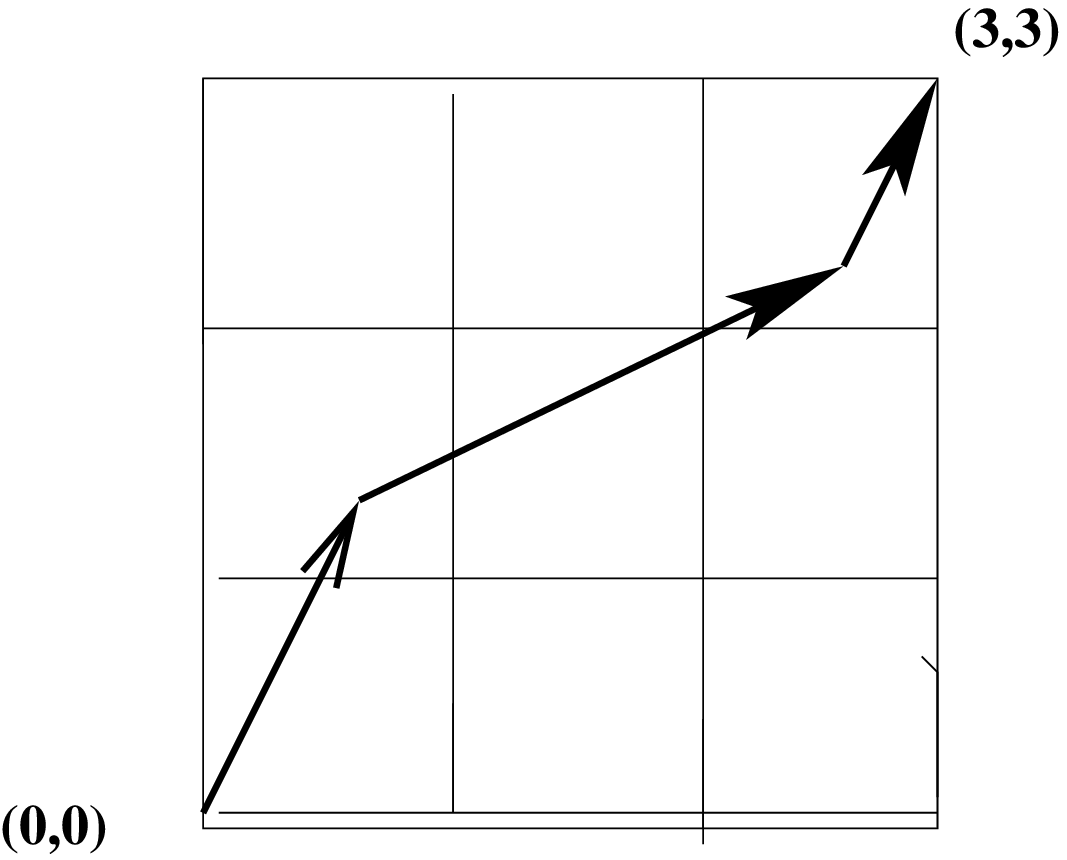}
\hspace{0.5cm}
\includegraphics[width=2cm]{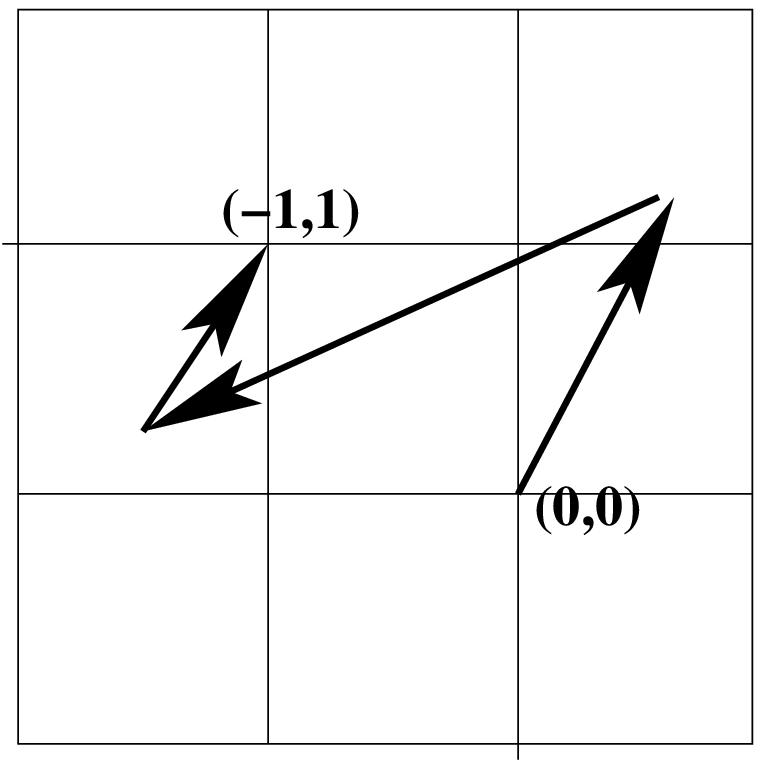}
\vspace{.5cm}
\noindent (c)~$R=(\frac{1}{3},\frac{2}{3}), p_1 = (1,2), p_2 = (2,1)$
\hfill
\includegraphics[width=2.5cm]{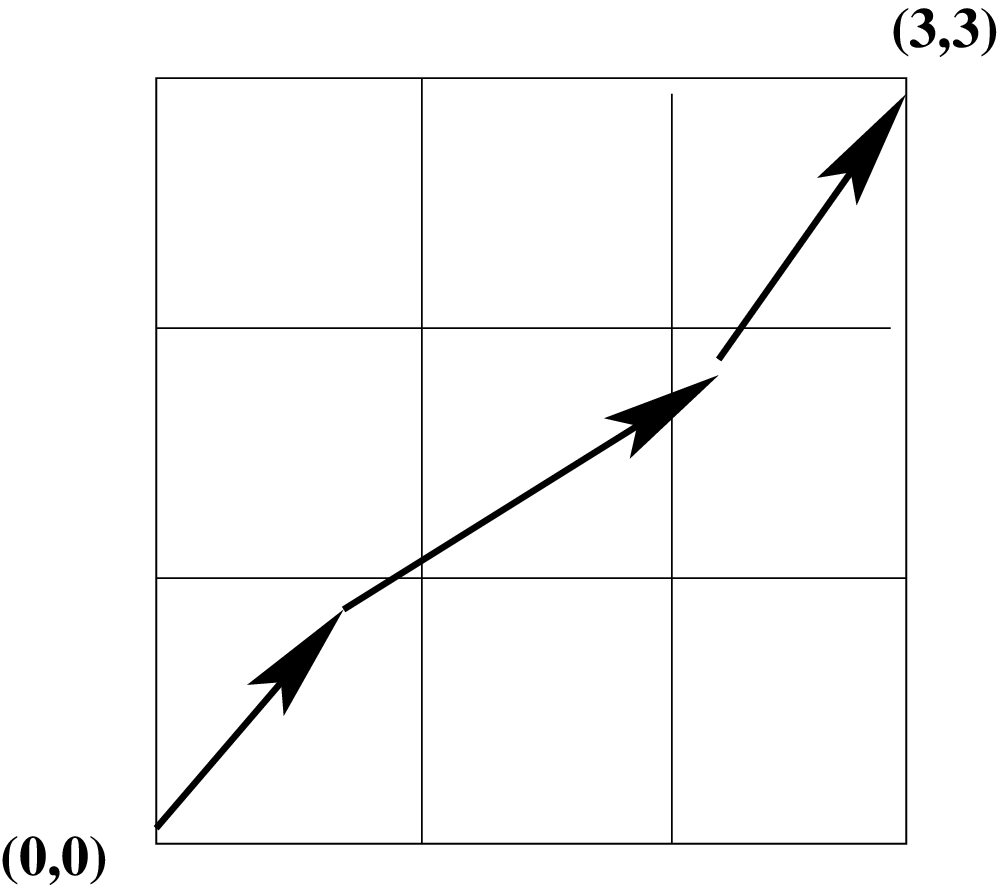}
\hspace{0.5cm}
\includegraphics[width=2cm]{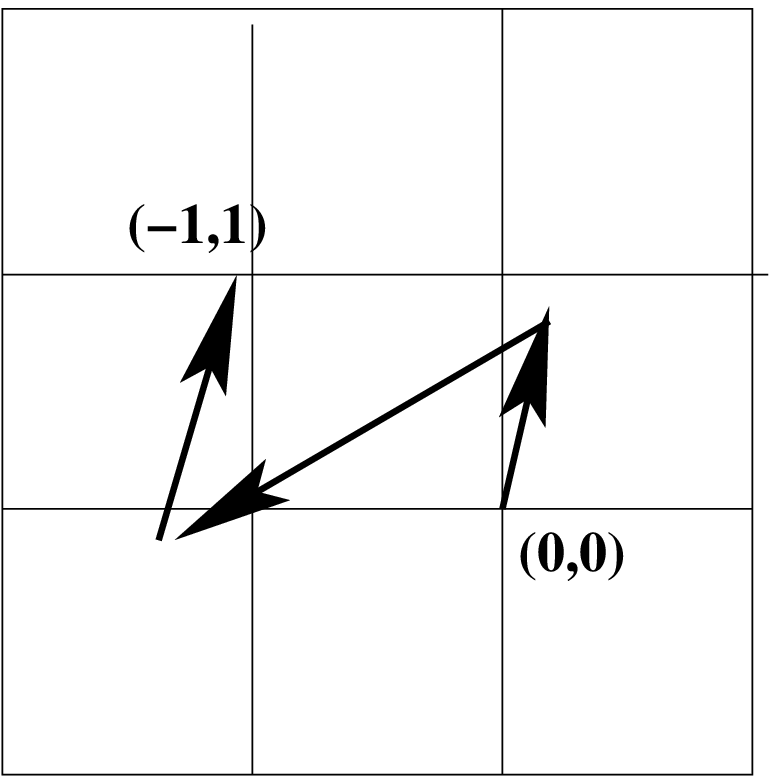}

\end{enumerate}

In each of the above examples, and for each intersection point $P, Q, R$, it
is clear that $p_1Pp_2\sim (m+s,n+t)$ and $p_1Pp_2^{-1}\sim(m-s,n-t)$.

To return to the bracket \rref{gold}, we assign classical functions to the straight paths $(m,n)$ as follows
\be
T(m,n) = e^{mr_1 + nr_2} + e^{- mr_1 - nr_2},
\label{tch}
\ee
i.e. $T(m,n) = {\rm tr}~~ U_{(m,n)}$ where $U_{(m,n)}$ is of the form
\rref{phi} with $r_1,\,r_2$ classical parameters. Setting $\{r_1,r_2\} = 1$ it follows 
that the Poisson bracket between these functions for two paths $(m,n)$ and $(s,t)$ is
\be
\{T(m,n), T(s,t)\} = (mt-ns)(T(m+s,n+t) - T(m-s,n-t))\{r_1,r_2\}
\label{pbt}
\ee
Equation \rref{pbt} may be regarded as a particular case of the Goldman
bracket \rref{gold} (up to setting $\{r_1,r_2\} = 1$),  since $(m,n)$
and $(s,t)$ have total intersection index $mt-ns$, and the rerouted paths $p_1Qp_2$
and $p_1Qp_2^{-1}$, where $p_1=(m,n)$ and $p_2=(s,t)$, are all
homotopic to $(m+s,n+t)$ and $(m-s,n-t)$ respectively.

The bracket \rref{pbt} is easily quantized using the triangle identity (see \rref{tri})
\be
e^{mr_1 + nr_2} e^{sr_1 + tr_2} = q^{(mt-ns)/2} e^{(m+s)r_1 + (n+t)r_2}
\ee
and the result is the commutator
\be
[ T(m,n), T(s,t)] = (q^{\frac{(mt-ns)}{2}} - q^{-\frac{(mt-ns)}{2}})
(T(m+s,n+t) - T(m-s,n-t)).
\label{tcomm} 
\ee
The antisymmetry of \rref{tcomm} is evident (from \rref{tch} 
$T(m,n)=T(-m,-n)$). It can be checked that \rref{tcomm} satisfies the Jacobi 
identity, and that the classical limit, namely $q \to 1, \hbar \to 0$, of the 
commutator \rref{tcomm}, given by

$$
\{,\} = lim_{\hbar \to 0}\frac{[,]}{i \hbar}
$$
is precisely \rref{pbt}.

Alternatively, there is a different version of equation \rref{tcomm} which treats 
each intersection point individually, and uses rerouted paths homotopic to 
``straight line'' paths as discussed previously, since we have 
already seen in Section \ref{hom} that homotopic paths no longer have the
same quantum matrix assigned to them, but only the same  matrix up to a
phase. Thus for an arbitrary PL  path $p$ from
$(0,0)$ to $(m,n)$, set
\be
T(p) = q^{S(p,(m,n))} T(m,n).
\label{tfactor}
\ee
The factor appearing in \rref{tfactor} is the same as that relating the 
quantum matrices $U_p$ and $U_{(m,n)}$, where $(m,n)$ is the straight path. 

\begin{figure}[hbpt]
\centering
\includegraphics[height=3cm]{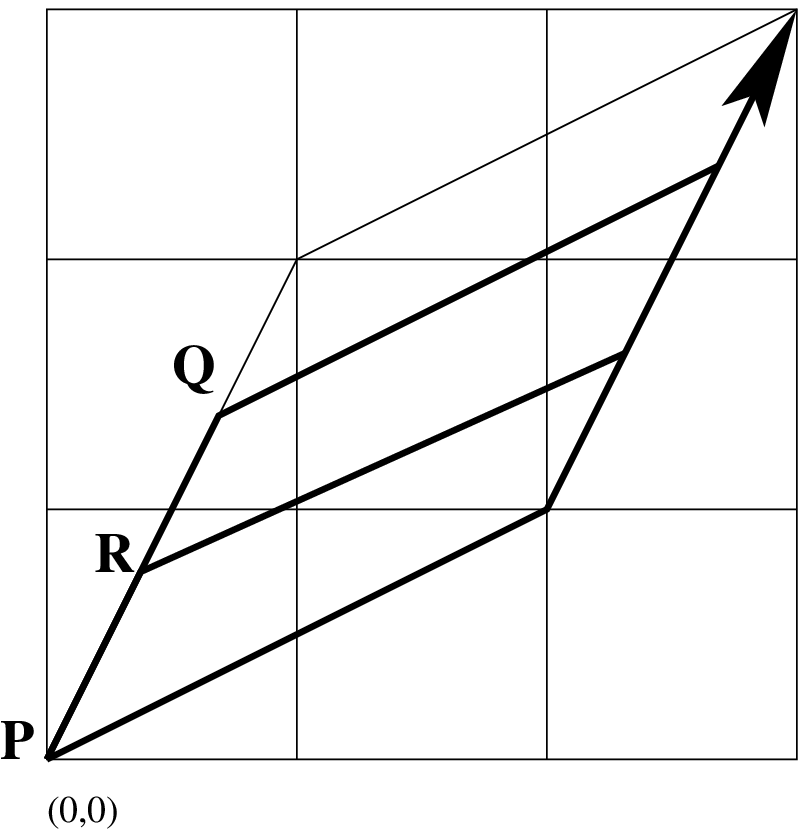}
\hspace{2cm}
\includegraphics[height=3cm]{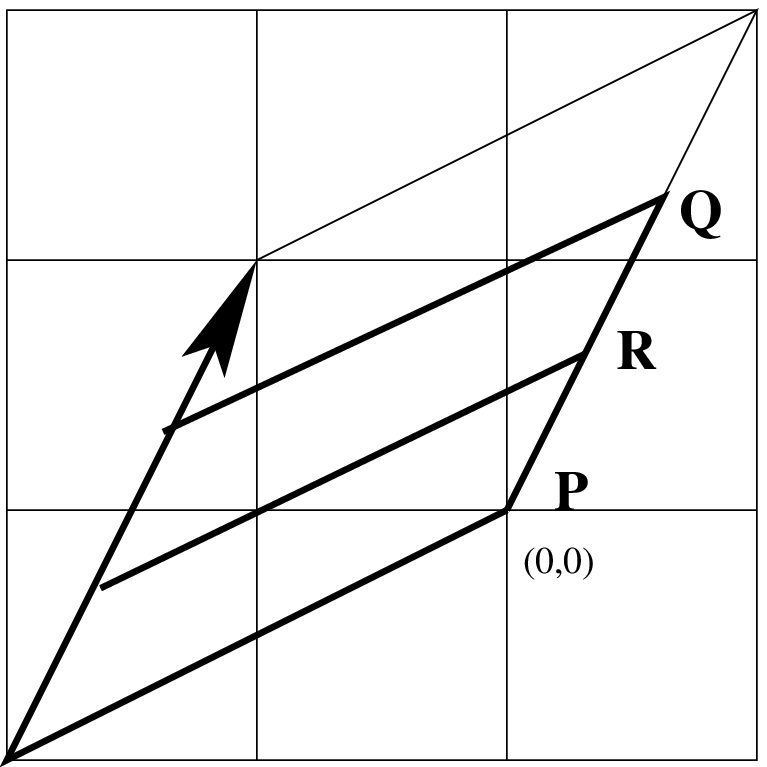}
\caption{ The reroutings $(1,2)S(2,1)$ and $(1,2)S(-2,-1)$ for $S=P,R,Q$ }
\label{p20a}
\end{figure}

We will show how to rewrite \rref{tcomm} in terms of the rerouted
paths for example 3, i.e.  $p_1=(1,2),\,p_2=(2,1)$. From \rref{tcomm}
\be
[ T(1,2), T(2,1)] = (q^{-3/2} - q^{3/2})(T(3,3) - T(-1,1)).
\label{tcommexp} 
\ee 
The intersections occur at the points $P, R, Q$ (in that order,
counting along $p_1$) as shown in Figure \ref{p20a}. For the positively
rerouted paths we have 
\bea
T((1,2)P(2,1)) & = & T((2,1)(1,2))=q^{3/2} T(3,3) \label{P}\\
T((1,2)R(2,1))& = & q^{-1}T((1,2)P(2,1)) \label{R}\\
T((1,2)Q(2,1)) & = & q^{-1} T((1,2)R(2,1)) \label{Q}
\end{eqnarray}
and for the negative reroutings
\bea
T((1,2)P(-2,-1)) & = & T((-2,-1)(1,2))=q^{-3/2} T(-1,1) \label{P-}\\
T((1,2)R(-2,-1))& = & q T((1,2)P(-2,-1)) \label{R-}\\
T((1,2)Q(-2,-1)) & = & q T((1,2)R(-2,-1)) \label{Q-}.
\end{eqnarray}

\begin{figure}[hbtp]
\centering
\includegraphics[height=3cm]{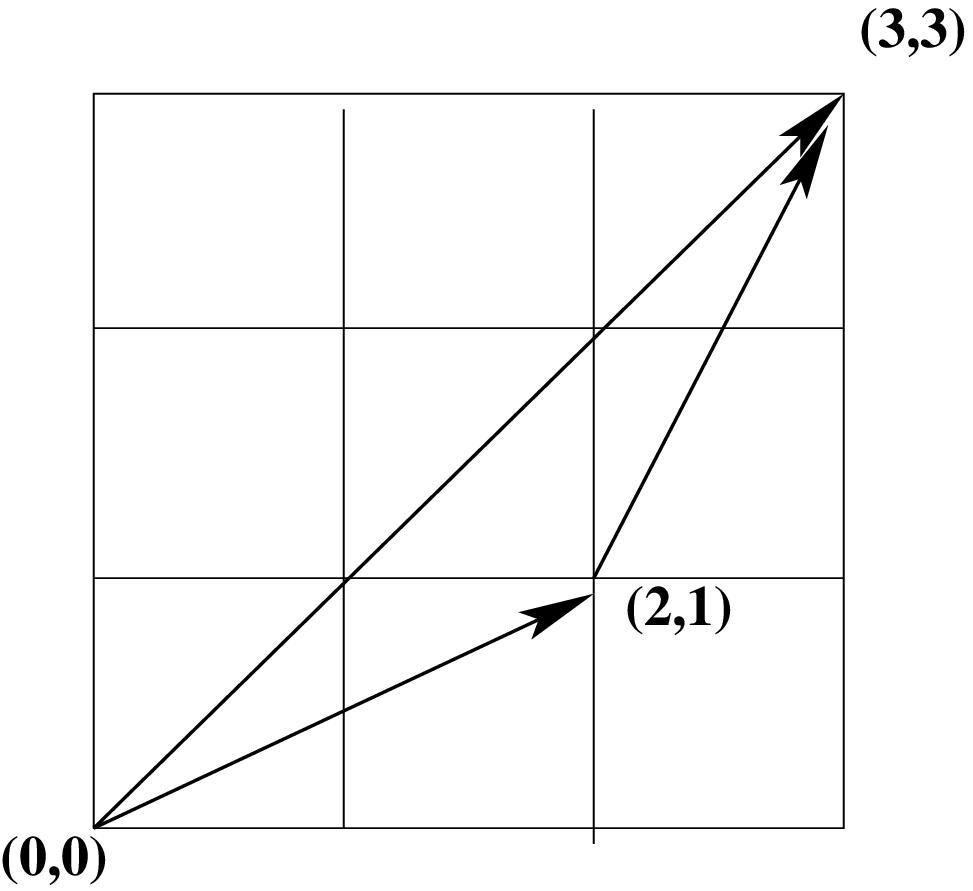}
\hspace{2cm}
\includegraphics[height=3cm]{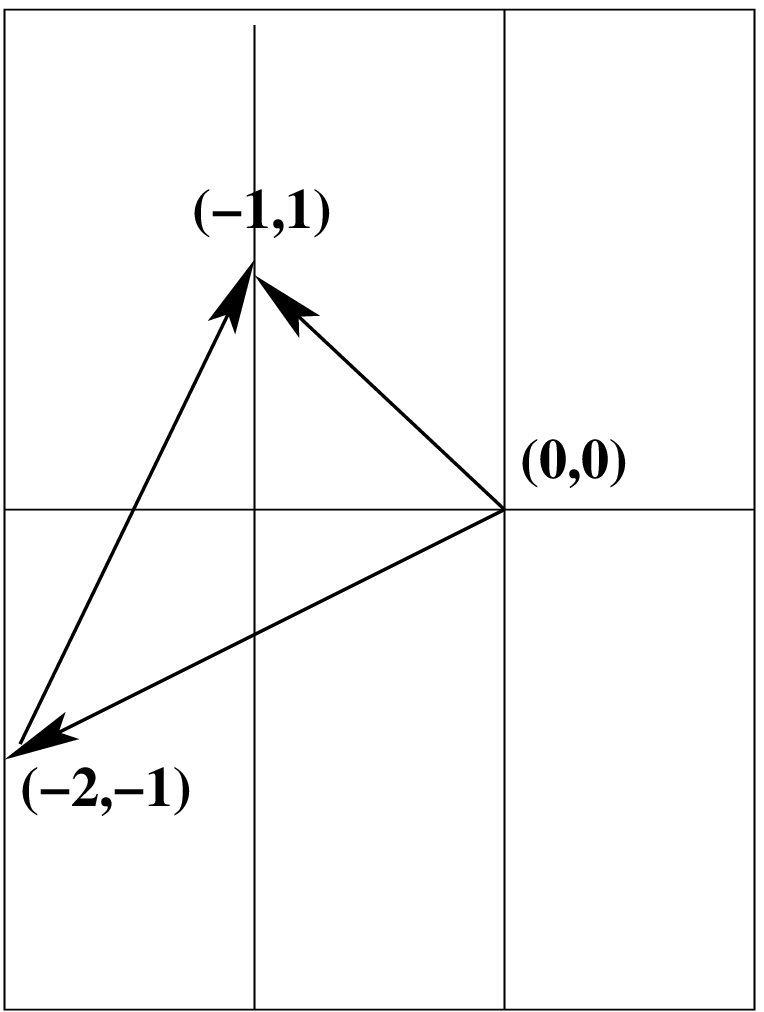}
\caption{Factors for the reroutings $(1,2)P(2,1)$ and $(1,2)P(-2,-1)$}
\label{trigs}
\end{figure}

The factors appearing in equations \rref{R}, \rref{Q}, \rref{R-} and \rref{Q-}
(the reroutings at $R$ and $Q$) are shown in Figure \ref{p20a}, 
where it is clear that each large parallelogram is divided into
three equal parallelograms, each of unit area. The factors in \rref{P} and
\rref{P-} (the reroutings at $P$) come from the triangle equation
\rref{tri}, and are shown in Figure \ref{trigs}, where the triangles have
signed area $+\frac{3}{2}$ and $-\frac{3}{2}$ respectively.

Now equation \rref{tcommexp} can be rewritten in the form:
\be
[T(1,2),T(2,1)] =\sum_{S=P,R,Q} 
(q^{-1} -1) T((1,2)S(2,1)) + (q-1) T((1,2)S(-2,-1)). 
\label{qgoldexp}
\ee

In the general case, for $p_1=(m,n)$ and $p_2=(s,t)$ with $mt-ns\neq 0$, 
we postulate that
\be
[T(p_1), T(p_2)] = \sum_{ Q \in p_1 \sharp p_2}
(q^{\epsilon(p_1,p_2,Q)} - 1)T(p_1Qp_2)  
+ (q^{-\epsilon(p_1,p_2,Q)} - 1)T(p_1Qp_2^{-1})
\label{qgold}
\ee
quantizes the Goldman bracket \rref{gold}.

\begin{figure}[hbpt]
\centering
\includegraphics[height=4cm]{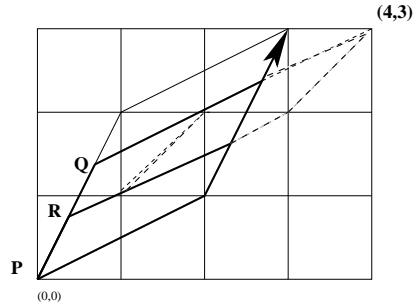}
\caption{ The area of the middle parallelogram is $1$}
\label{pp20a}
\end{figure}

We have proved equation \rref{qgold} as follows: first assume that both $p_1$ and $p_2$
are irreducible, i.e. not multiples of other integer paths, and study the reroutings 
$p_1 Qp_2$ at $Q$. They are paths similar to those
of Figure \ref{p20a}, namely  following $p_1$ to $Q$, then rerouting along a
path {\it parallel} to $p_2$, then finishing along a path {\it parallel} to
$p_1$. The reroutings along $p_2$ must clearly pass through an integer
point inside the parallelogram formed by $p_1$ and $p_2$ (apart from when
the intersection point is the origin). They also clearly pass through only one
integer point since $p_2$ is irreducible. Consider two adjacent lines inside
the parallelogram parallel to $p_2$ and passing through integer points. 
The area of each parallelogram between them is $1$.
Consider for instance one of the middle parallelograms in 
Figure \ref{p20a} (whose area we saw previously was $1$ as the three 
parallelograms are clearly of equal area and the area of the large 
parallelogram is $3$). 
This is the same area as that of a parallelogram with vertices at
integer points, as can be shown, for example, by cutting it into two pieces 
along the line between (1,1) and (2,2), then regluing them together into a 
parallelogram with vertices at (1,1), (2,2), (3,2) and (4,3), as indicated in
Figure \ref{pp20a}. This latter area is equal to $1$ from Pick's theorem
\cite{pick} which states that the area $A(P)$ of a lattice polygon $P$ is
\be A(P) = I(P) + B(P)/2 - 1, \label{pick} 
\ee 
where $I(P)$ is the number of interior lattice points and $B(P)$ is the number
of boundary points 
(for the parallelogram in the example $I(P)=0$ since
the lines parallel to $p_2$ are adjacent, and $B(P)=4$ from the integer
points at the four vertices, so $A(P) = 0+4/2-1=1$.) Therefore in general the
parallelogram determined by $p_1$ and $p_2$, whose total area is $A=
|mt-ns|$, is divided up into $A$ smaller parallelograms of equal area by lines
parallel to $p_2$ passing through the interior integer points of the
parallelogram. The fact that the total area is equal to the number of internal
integer points $+1$ is again a consequence of Pick's theorem. 

We can now calculate the first term (the positive reroutings shown for the
example in Figure \ref{p20a}) in the sum on the r.h.s. of
\rref{qgold},  using equation \rref{tfactor}, and show that it is equal to the
first term on the  r.h.s. of \rref{tcomm}. Consider first the case
$\epsilon(p_1,p_2,Q) = -1$. The  rerouting at the origin satisfies, using
the triangle equation \rref{tri}, 
\[
T(p_1\, (0,0)\, p_2) = q^{A/2} T(m+s,n+t),
\]
where the area of the parallelogram determined by $p_1,\,p_2$ is $A=-(mt-ns)$.  
The next rerouted path adjacent to $p_1\, (0,0)\, p_2$, rerouted at $Q_1$ say 
(in the example $Q_1=R$) satisfies 
\[
T(p_1\, Q_1\, p_2) = q^{-1}T(p_1\, (0,0)\, p_2)
\]
since we have shown that the signed area between the paths is $-1$. Similarly 
each successive adjacent path rerouted at $Q_2, Q_3, \dots$ satisfies
\be
T(p_1\, Q_i\, p_2) = q^{-1}T(p_1\, Q_{i-1}\, p_2).
\ee
with $Q_0$ the origin $(0,0)$. It follows that 
\begin{eqnarray}
\lefteqn{\sum_{Q \in p_1 \sharp p_2}(q^{-1}-1)T(p_1 Q p_2)}\nonumber \\ 
&=& (q^{-1}-1)q^{A/2}(1+q^{-1} + \dots + q^{-(A-1)})T(m+s,n+t) \nonumber \\
&=& (q^{-1}-1)q^{A/2}\frac{1-q^{-A}}{1-q^{-1}}T(m+s,n+t)\nonumber \\
&=& (q^{-A/2} - q^{A/2})T(m+s,n+t)\nonumber \\
&=& (q^{(mt-ns)/2} - q^{-(mt-ns)/2})T(m+s,n+t).
\label{ep+}
\end{eqnarray}
When $\epsilon(p_1,p_2,Q)=+1$ the calculation is identical to \rref{ep+} 
but with $q$ rather than $q^{-1}$, 
and with the area of the triangle now equal to $A/2$, where $A=mt-ns$, 
namely
\begin{eqnarray}
\lefteqn{\sum_{ Q \in p_1 \sharp p_2}(q-1) T(p_1\, Q\, p_2)} \nonumber \\
&=& (q-1)q^{-A/2}(1+q^{1} + \dots + q^{A-1}) T(m+s,n+t)\nonumber \\
&=& (q^{A/2} - q^{-A/2}) T(m+s,n+t)\nonumber \\
&=& (q^{(mt-ns)/2} - q^{-(mt-ns)/2}) T(m+s,n+t).
\label{ep-}
\end{eqnarray}

Diagrammatically this corresponds to dividing up the first parallelogram in Figure
\ref{p20a} by lines passing through the integer points in the interior, but
parallel to $(1,2)$, as opposed to $(2,1)$.

In an entirely analogous way the second terms (the negative reroutings) on the 
r.h.s. of \rref{tcomm} and \rref{qgold} can be shown to be equal - the second
figure of Figure \ref{p20a} can be used as a guide\footnote {The antisymmetry
of \rref{qgold} can be checked for our example $p_1=(1,2),p_2=(2,1)$, both
irreducible, by noting that the intersections  occur at the same points (but in
a different order, namely  $P, Q, R$).}.

When $p_1$ is reducible, i.e. $p_1 =c(m',n'), \, c\in \mathbb{N}, m', n' \in
\mathbb{Z}$, and $p_2$ is irreducible, formula \rref{qgold} applies exactly as
for the irreducible case, since there are $c$ times as many rerouted paths
compared to the case when $p_1=(m',n')$. An example is $p_1=(2,0),\,
p_2=(1,2)$, where the first term on the r.h.s. of \rref{tcomm} is equal to the 
first term on the r.h.s. of \rref{qgold}:
\begin{eqnarray}
(q^2-q^{-2}) T(3,2) & = & (q-1)q^{-2} (1+q+q^2 + q^3) T(3,2)\nonumber \\
&=& \sum_{ Q \in p_1 \sharp p_2} (q-1) T(p_1Q p_2).
\end{eqnarray}
There are four rerouted paths in the final summation, rerouting at $(0,0)$,
$(1/2,0)$, $(1,0)$ and  $(3/2,0)$ along $p_1$.

If $p_2$ is reducible we must use multiple intersection numbers in \rref{qgold},
 i.e. not simply $\pm 1$. Suppose $p_1=(m,n)$ and $p_2=(s,t) = c(s',t')$, $
c\in \mathbb{N}, s', t' \in \mathbb{Z}$. Then for example the first
term on the r.h.s. of \rref{qgold} with $mt-ns>0$ is 
\begin{eqnarray}
\lefteqn{(q^{(mt-ns)/2} - q^{-(mt-ns)/2}) T(m+s,n+t)} \nonumber \\
&=& (q^{c(mt'-ns')}- 1) q^{-(mt-ns)/2} T(m+s,n+t) \nonumber\\
&=& (q^c-1) q^{-(mt-ns)/2} (1 + q^c + \dots + q^{c(mt'-ns'-1)}) T(m+s,n+t)\nonumber \\
&=& \sum_{ Q \in p_1 \sharp p_2} (q^c-1)  T(p_1Q p_2).
\label{multint}
\end{eqnarray}

The factor $(q^c-1)$ is the {\it quantum} multiple intersection number at the
$mt'-ns'$ intersection points. The calculation can be regarded as doing equation \rref{ep-} backwards and substituting $q$ by $q^c$ and $mt-ns$ by $mt'-ns'$. An example is 
$p_1=(2,1),\, p_2= (0,2)$, for which double intersections occur along $p_1$ at the origin
and at $(1,1/2)$. From \rref{multint} the first term on the r.h.s. of \rref{qgold} is

\begin{eqnarray}
(q^2-q^{-2}) T(2,3) & = & (q^2-1)q^{-2} (1+q^2) T(2,3) \nonumber \\
&=& (q^2-1) (T(p_1\, (0,0)\, p_2) + T(p_1\, (1,1/2)\, p_2).
\end{eqnarray}

\section{Conclusions}

There are some surprising features of the quantum geometry that emerge from the use of a constant quantum connection. The phase factor appearing in the fundamental relation \rref{fund2} has a geometrical origin as the signed area phase relating two integer PL paths, corresponding to two different loops on the torus. This leads to a natural concept of $q$-deformed surface group representations. It follows that the classical correspondence between flat connections (local geometry) and holonomies, i.e. group homomorphisms from $\pi_1$ to $G$ (non-local geometry) has a natural quantum counterpart. 

The signed area phases also appear in a quantum version \rref{qgold} of a
classical bracket \rref{gold} due to Goldman Ref.~\refcite{gol}, where
classical intersection numbers $\pm \ep(p_1,p_2,Q)$ are replaced by quantum
single and multiple intersection numbers $(q^{\pm \ep(p_1,p_2,Q)}-1)$. 

The quantum bracket for homotopy classes represented by straight lines \rref{tcomm} is easily checked since all the reroutings are homotopic. However the r.h.s. of the bracket 
\rref{qgold} may be expressed in terms of rerouted paths using the signed area phases and a far subtler picture emerges. 

It is not difficult to show that the Jacobi identity holds for the commutator for straight 
paths \rref{tcomm} since the r.h.s. may also be expressed in terms of straight paths, with suitable phases. It must also hold for \rref{qgold} since they are equivalent. We have checked it explicitly for a number of arbitrary PL paths, without identifying homotopic paths. 

It should also be possible to treat higher genus surfaces (of genus $g$) in a similar 
fashion by introducing the same constant quantum connection on a domain in the 
$xy$ plane bounded by a $4g$--gon with the edges suitably identified \cite{hil}.  
One could then define holonomies of PL loops on this domain and study their behaviour under intersections, as studied here for $g=1$. In fact this treatment is ideal for 
$g>1$ since there intersections $\ep \ge 2$ are necessary (see e.g. Ref.~ \refcite{NR1}).

\section*{Acknowledgments}

This work was supported by the Istituto Nazionale di Fisica Nucleare
(INFN) of Italy, Iniziativa Specifica FI41, the Italian Ministero
dell'Universit\`a e della Ricerca Scientifica e Tecnologica (MIUR), and the {\em Programa Operacional Ci\^{e}ncia e Inova\c{c}\~{a}o 2010}, project number 
POCI/MAT/60352/2004, financed by the {\em Funda\c{c}\~{a}o para a 
Ci\^{e}ncia e a Tecnologia} (FCT) and cofinanced by the European 
Community fund FEDER.


\end{document}